\documentstyle[aps,preprint,psfig]{revtex}
\begin{document}
\title{\bf Thermodynamic Theory
of Weakly Excited Granular Systems}
\author{Hisao Hayakawa$^{1}
$\thanks{E-mail address: hisao@phys.h.kyoto-u.ac.jp}
and Daniel C. Hong$^2$\thanks{
E-mail address: dh09@lehigh.edu}}
\address{$^{1}
$Graduate School of Human and Environmental Studies,
Kyoto University, Yoshida, Sakyo, Kyoto 606-01, Japan \\
$^{2}$ Department of Physics, Lewis Laboratory,
Lehigh University, Bethlehem, Pennsylvania 18015}
\maketitle
\begin{abstract}
We present a thermodynamic theory of weakly excited two-dimensional
granular systems
from the view point of elementary excitations of spinless Fermion systems.
We introduce a global temperature $T$ that is associated
with the acceleration amplitude $\Gamma$ in a vibrating bed.
We show
that the configurational statistics of weakly excited
granular materials in a vibrating bed obey the Fermi statistics.
\end{abstract}
\noindent P.A.C.S. numbers 81.35.+k,46.10.+z,05.70.-a,05.20.Dd
\vskip 0.2 true cm
Granular system is robust to thermal disturbances because its entity is a
macroscopic object.\cite {1}
  For this reason, the granular system is effectively in
the ground state at any finite temperature and the excitation 
may be achieved by subjecting the system to vibration or shaking.  Such an
external stimulus will inject energy at a constant rate but the energy will
be dissipated via collisions, leading the system to reach a steady state. 
Dynamics of such a steady state are quite
complex, where convection \cite {2}, density waves \cite {3}, segregation
\cite {4}, anomalous sound propagation \cite {4b}
 and even turbulent behaviors \cite {5}
have been observed.

There are some indications that fluctuations in
physical quantities of granular systems persist over the size of the system
\cite {7}
and intrinsically nonequilibrium clustering instabilities  appear for
particles with large coefficient of restitution \cite {8}.
In such cases, we may eventually
have to question the validity of the hydrodynamics\cite{hydro}
with the aid of kinetic theory\cite{kin},
though some attempts have been made to capture some of the
essential features of granular convections
based on phenomenological hydrodynamics models\cite{hyh}.

In spite of the above negative signs,
the validity of the thermodynamics concept
has been suggested by several theoretical papers\cite{9,hjh}
and experimental papers\cite{6,MD,Warr,knight}.
In particular, Knight et al\cite{knight} have observed a logarithmic
relaxation in compaction processes in a three dimensional vibrating bed,
 which can be understood as
the consecutive transitions among the metastable (glassy) configurations.
This suggests the validity of the free volume (or hole) theory
\cite{hill} used for the dense liquid theory as will be shown later.
In two dimensions, in particular,
the situation  is much simpler than 
in three dimensions,
because the particles can form a lattice structure without glassy
configurations.   For example,
the experiment by Clement and Rajchenbach \cite {6}
has suggested that nontrivial
and distinctive {\it configurational} statistics appear to exist for
excited granular systems in a vibrating bed.
The experiment was conducted with steel balls that have small coefficient
of restitution and was monitored carefully to suppress the convection
with a suitable choice of the boundary condition.  They
then observed that the ensemble-averaged density profile 
obeys a universal function that is independent
of the phase of oscillations.
The experimental result in Ref.\cite{6} has been recovered by
a simulation based on the distinct element method\cite{MD} and
has been generalized to the case of strong excitions.\cite{Warr}
The existence of such a distinctive configurational statistics, which resembles
the problem of packing, 
appears to be a fairly convincing evidence that kinetic aspects 
of the vibrating bed might have been decoupled 
from the statistical configurations averaged over many
ensembles and time sequences.
\vskip 0.2 true cm
Such a simple observation in two dimensional weak dissipation cases
enables one to make some progress in characterizing the excitation
of vibrating beds in two ways: first,
if the kinetics is indeed separated out, then the configurational properties
should be determined by the principle of maximum entropy or equivalently
the minimization of free energy.  Second,
the validity of such variational methods should
 be carefully
checked against experiments.
The test may lead to
further conceptual
advances, establishing the fact that
a weakly
driven nonequilibrium dissipative vibrating bed
with the vibration intensity $\Gamma$
may be viewed as
a thermodynamic equilibrium state at a finite temperature $T$
in the near elastic limit,
{\it if} one's focus is exclusively on the configurational properties.
The precise relation between $\Gamma$ and $T$, however, 
has yet to be determined.
The purpose of this Letter is to advance such a simple
observation into systematic investigations and to
formulate a thermodynamic theory of powders, at least,
in two dimensional systems, from the view point of
elementary excitations such as
the Fermi liquid theory\cite {6b} in condensed matter physics.
Our formulation 
may open a way to visualize
the invisible quantum behaviors of Fermions or the microscopic behaviors of
dense fluids through the manipulation of
granular materials.
\vskip 0.2 true cm
The starting point of
our thermodynamic formulation\cite{Korea}
is the recognition that
granular state in a vibrating bed is an excited state and the degree of the
excitation is controlled by the global parameter $\Gamma$.
Since we are concerned here with the configurational property of such a system,
it is natural to associate a similar {\it global}
temperature $T$, but a care must be
taken here because the global temperature $T$ must have a well defined
thermodynamic meaning.  One of the essential requirement is that
$T$ must satisfy the
statistical definition of the thermodynamic temperature, namely:
$ T =\partial U/\partial S$,
where $U$ is the total energy and $S$ is the entropy of the system.
Notice that the conventional kinetic temperature, which is in general
a local function, is not identical to the thermodynamic temperature.
In fact, $T$ can be nonzero without kinetic energy, because $U$
contains the potential which is a function of the entropy.
When the contribution from the kinetic energy
is much smaller than the potential energy,
the global temperature may be more appropriate than the local granular
temperature to characterize the state of granules as the idea used
in the free volume theory\cite{hill}.
One can easily show that this parameter $T$ is almost
identical with the
compactivity $X$ defined through the free volume
introduced by Edwards and his coworkers. \cite {9}
While the
compactivity $X$ has never been computed nor related in any manner with the
experimental control parameters such as $\Gamma$, our formulation
will enable us to determine the
explicit relation between $T$ and $\Gamma$.
To be more specific, we first view the system of granular particles
as the lattice gas, which  
can be regarded as the simplest version of
the free volume (hole) theory\cite{hill}.
We now assign virtual lattice points by
dividing the vibrating bed of width $L$ and the height $\mu D$
into cells of $D\times D$ with  the diameter of the grain $D$.
  Each row, $i$, is
then associated with the potential energy
$\epsilon_i=mgz_i$ with $z_i=(i-1/2)$D
and $m$ the mass of the grain.  
The degeneracy, $\Omega$, of
each row is 
$\Omega=L/D$.
For a weakly excited system with $\Gamma \approx 1$,
the kinetic energy may be neglected and the potential energy
dominates, for which case
the most probable configuration should be determined by
the state that maximizes the entropy in the microcanonical ensemble approach.
\vskip 0.3 true cm
The entropy, $S$,
is defined as $S= \ln W$ with $W$
the total number of ways of distributing $N$ particles
into a system.
Excluded volume interactions
do not allow two grains to occupy the same states and thus
the statistics is given by the Fermi statistics.
We find:
\begin{equation} W =\Pi_i[\Omega!/N_i!(\Omega-N_i)!]. \label {(1)}
\end{equation}
We now maximize $S$ with
constraints that characterize the system, namely the fixed number of particles
$N$ and the mean steady state system energy
$<U(T(\Gamma))>$,
namely
\begin{equation}\label{(2)} \sum_i N_i =N ,\quad
\sum_i N_i\epsilon_i = U . \end{equation}
The maximization of $S$ then
yields that the density profile, $\phi(z)$, which is the
average number of occupied cells at a given energy level, must
be given by the Fermi distribution:
\begin{equation} \phi(z)=N_i/\Omega = 1/[1+\exp(\beta(z-\mu))] , \label{(3)}
\end{equation}
where $\beta\to mgD/T$ in the low temperature limit,
 the height $z=z_i/D$ and the Fermi
energy $\mu$ measured
in units of $D$ is
the initial number of layers.
Note that both z and $\mu$ are measured from the bottom layer.
The Fermi analogy is valid when 
$\mu\gg n_l$
with $n_l$ is the number of fluidized layer.  For a
non-interacting electron gas, this ratio is of order $10$
to $10^2$.  Now, since
the injected energy,$ E_i$,
to the system is of order $mA^2\omega^2/2$ and
the potential energy, $E_p$,
to fluidize particles on the top $n_l$ layers is of order $mgn_lD$,
by equating the two,
 we find a necessary condition for the fluidization of the top $n_l$
layers, namely: $n_l\sim \Gamma (A/2D)$.  
For $\Gamma\sim 1$, 
the Fermi statistics will be valid,
if $\mu \gg A/D$. 
\vskip 0.2 true cm
Our only remaining task is then
to relate the temperature $T$ to the control parameter
$\Gamma$.  Here we do not calculate the entropy and the energy directly,
because nonequilbrium and kinetic  characteristics may appear in the relation
between the temperature $T$ and $\Gamma$.
Since we have determined
the density profile, we find from (2) the energy {\it per} particle:
\begin{equation}
\bar u(T)={\frac{U(T)}{N}} = {\frac{mgD\mu^2}{2}}[1+{\frac{\pi^2}{3}}
({\frac{T}{mgD\mu}})^2] +\cdots \label{(4)} , \end{equation}
where the first term is the ground state energy
and the second term is the increase in energy due to {\it thermal}
 expansion, which results in
 the shift in the center of mass, $\bar h(T) = \bar u(T)/mg $.
\begin{equation}
 \bar h(T) = h(0)[1+{\frac{\pi^2}{3}}({\frac{T}{mgD\mu}})^2]
+\cdots \label{(5)}
\end{equation}
with $h(0)=D\mu^2/2$.  We now make a crucial observation that
for a weakly excited granular system, most excitations occur near the
Fermi surface, which may be effectively well
represented by the motion of a {\it single}
particle on the Fermi surface that is in contact with the vibrating plate.
If the maximum height that a single ball bouncing in a vibrating plate with
the intensity $\Gamma$ is denoted by $H_0(\Gamma)$, then $H_0(\Gamma)$ is
determined by the equation that
describes the trajectory of a single ball on a vibrating
plane with the
intensity $\Gamma$. \cite{hyh,10}  The relative distance, $\Delta(t)$,
 between the ball
and the vibrating plate is given by:
\begin{equation}
 \Delta (t) = \Gamma(\sin(t_0)-\sin(t)) + \Gamma \cos(t_0)(t-t_0) -
{\frac{1}{2}}(t-t_0)^2\label{(6)}
\end{equation}
in units of $g=\omega=1$,where $t_0=\sin^{-1}(1/\Gamma)$.
The maximum, $H_0(\Gamma)$, can be obtained from (\ref{(6)})
numerically and it
 is effectively equivalent to the expansion of
the volume due to {\it kinetics}.
Since the Fermi distribution near $T=0$ can be approximated by a piecewise
linear function and $H_0(\Gamma)$ is thought to be the edge of the
function,  we expect
$H_0(\Gamma) \approx \Delta h/2
=(\bar h(T) - h(0))/2$.  By equating the thermal expansion, (5),to the kinetic
expansion, $H_0(\Gamma)g/\omega^2$ in physical units, we now complete our
thermodynamic
formulation by presenting the explicit relation between $T$ and 
$\Gamma=A\omega^2/g$:
\begin{equation}
 T = {\frac{mg}{\pi\omega}}(3Dg H_0(\Gamma))^{1/2} .  \label{(7)}
\end{equation}
We point out that the energy is an extensive quantity along the
horizontal direction, but is not in vertical direction where strong anisotropy
is present due to gravity.
Further, $T$ has a gap at $T=0$ because the time between the launching
and landing of the ball is always finite for $\Gamma>1$.
 Figure 1  shows the fitting of the
experimental density profile for $\Gamma=4$ of Clement and Rajchenbach
\cite {6} by the scaled Fermi distribution, $\phi(y)=\rho(y)/\rho_c$,
with $\rho_c\approx 0.92$ the closed packing density for the hexagonal packing.
The fitting value of $T/mg$ is $2.0 [mm]$ with $\mu D=30.5 [mm]$,
while eq.(7) yields $T/mg \approx 2.6 [mm] $ \cite{evaluation}.
The agreement between the two is fairly good in spite of
such a simple calculation. This expression also agrees with
the simulation result\cite{MD}.
Note that the
detailed expression of $H_0(\Gamma)$ depends on the manner by which the
grains are excited and we expect that our main scaling prediction
of eq.(\ref{(7)}), namely
$T\propto g^{3/2}D^{1/2}/\omega$, will hold even for systems driven not
by sinusoidal waves.
Next, it is well known that the
{\it specific heat} per particle, $C_v = d\bar u/dT$,
can be written as the fluctuations in the 
energy, namely $<(\Delta \bar u)^2>=<(\bar u(z)-\bar u)^2> =
T^2C_v$ and thus we predict
the scaling relation for
the fluctuations in potential energy, 
or in the center of mass,
\begin{equation}
\label{(8)}
<(\Delta \bar u)^2>={\frac{\pi^2}{3}}{\frac{T^3}{mgD}}
\propto g^{7/2}D^{1/2}\omega^{-3}.
\end{equation}
  Certainly,
more experiments or  simulations
would be desirable to test our theory.

\vskip 0.2 true cm
Three comments are in order: First,
in three dimensional systems the situations become far more complex and
our ideal Fermi description based on simple lattice gas picture
certainly requires modification, primarily for two reasons: first, holes
are not equivalent to particles and second, many metastable configurations
exist, which results in the hysteresis-dependent density profile as observed
in the experiment of Knight et al.\cite{knight}
Even in this case, however, lattice picture may be valid
because the mean free
path of the grains is of order of a few particle diameters and
the basic granular state is not a gas but a
crystal.  Hence, the
free energy approach adopted in this paper is more appropriate than the
kinetic theory in studying the granular state.
Such an approach is consistent with
the free volume theory of liquid state \cite{hill}, which assumes that
the dominant process involving particle rearrangement is a hopping 
with the rate determined by the activation energy $A$.
Within this picture, the probability
of the hopping of a particle from a position {\bf r}
is proportional to $\exp[-A({\bf r})/T]$ with 
$A({\bf r})  \simeq a\phi({\bf r})/(1-\phi({\bf r}))$.
The relevant time scale $\tau$ that determines the time evolution
of the compaction is then
given by
$\tau/\tau_0\simeq \exp[b \phi/(1-\phi)]$ with $b=a/T$, or
$\phi(t)\simeq \ln (t/\tau_0)/(b+\ln(t/\tau_0))$, where we have
replaced $\tau$ by $t$.  This is consistent with the experimental result
reported by Knight et al. \cite{knight}
Notice that such an activation dominated process does not have
smooth increase of the density.  
Although this kind of slow relaxation may
not be unexpected
even in two dimensional systems, the strong geometrical constraint may
suppress such a slow process.
We point out that attempts have been made to determine the
thermodynamic temperature defined in this paper
by measuring the effective viscosity for fluidized beds with a mixture of
gas and particles.  
The results \cite{ichiki} seem to support the validity
of the free volume theory.  This is an indirect
confirmation of the validity of
our free Fermion picture based on the free volume theory even
in three dimensional
systems.

Second, interactions among particles.
It is known that even for hard core particles, an effective attractive
interaction exists through the direct correlation function \cite{Hansen},
which will induce the curvature term.
In the presence of such a curvature term, while it may be difficult to
quantitatively compute the surface tension due to the spatial inhomogeneity,
it is nevertheless obvious that we  can
define surface tension for excited granular materials in a vibrating bed.
We may need more systematic experiments and careful theoretical argument
to resolve the question of surface tension in excited granular materials
\cite{2}.
\vskip 0.2 true cm
Third, the effect of dissipation. 
For a simple one-dimensional system of
$N(\gg 1)$ particles connected by springs, where the end particle
at the bottom is driven by an external sinusoidal force,
the equation of
motion for $n-$th bead is then given by:
$ m \ddot z_n+m\zeta \dot z_n-k\partial_n^2 z_n=m A \omega^2 \cos\omega t
\delta_{n,0}$.
For such a linear system, 
the effective amplitude $A_{eff}$ 
felt by the particle at the top
is expressed by $A_{eff}=A/\sqrt{1+(\zeta/\omega)^2}$, where the dissipation
constant
$\zeta=-(\omega/\pi) \ln e$ with the restitution constant $e$\cite{tag}.  
Thus, 
the correction for $A$ is very small. 

\vskip 0.2 true cm
\vskip 0.5 true cm
After we have submitted this Letter, we became aware of two
recent papers in which
thermodynamic concepts such as one developed in this paper
are useful in highly excited vertical vibration of
granular materials\cite{hansen} and in horizontal vibration of granular
materials\cite{ristow}.
In these papers, the authors
have demonstrated that stationary nonequilibrium state of the vibrating
bed exhibits similarities with the thermally equilibrated fluid, consistent
with the assumption of this paper, and
the spatial
correlation function (two-point correlation) can be an index of
solid-fluid transition of granular materials and can be approximated by
the equilibrium distribution function in the fluid phase.
\vskip 1.0 true cm
We would like to express our sincere gratitude to N.Goldenfeld
for his encouragement of
 this study during HH's stay at U. of Illinois and 
his collaboration in the early stage of this study,
Y. Oono, G. Baym, and H.J.Herrmann for
helpful discussions concerning the Fermi liquid theory, Ryan Mitchell for
graphical assistance, and 
J. A. McLennan on various aspects of the
kinetic theory of dense gases.
HH is, in part, supported by
Grant-in-Aid for Scientific Research from Japanese Ministry of Education,
Science and Culture (No. 08226206 and 08740308).

\vspace*{0.5cm}

\vfill\eject

\vfill\eject
\noindent {\bf Figure Caption}
\vskip 1.0 true cm
\noindent Fig.~1. Comparison between the experiment
and the theoretical prediction.  The circles are
the data by Clement and Rajchenbach(ref.14) and the dotted line is the
Fermi distribution function $\phi(z)$ (eq.(3)).

\newpage
\thispagestyle{empty}
\centerline{\hbox{
\psfig{figure=haya-hong-prl}
}}


\begin{references}
\bibitem{evaluation}
This good agreement may be a little surprising for strong excitation.
For $\Gamma=4$, $t_0=0.25268$, and the solution of $d\Delta(t)/dt=0$ is
given by $t_{max}=4.59483$, for which case $H_0(4)=12.3623$.
In ref.\cite{6} $D=3 [mm]$ and $\omega/(2\pi)=20 [Hz]$.
For $\Gamma=4$,
the bouncing solution becomes chaotic.
\end{references}
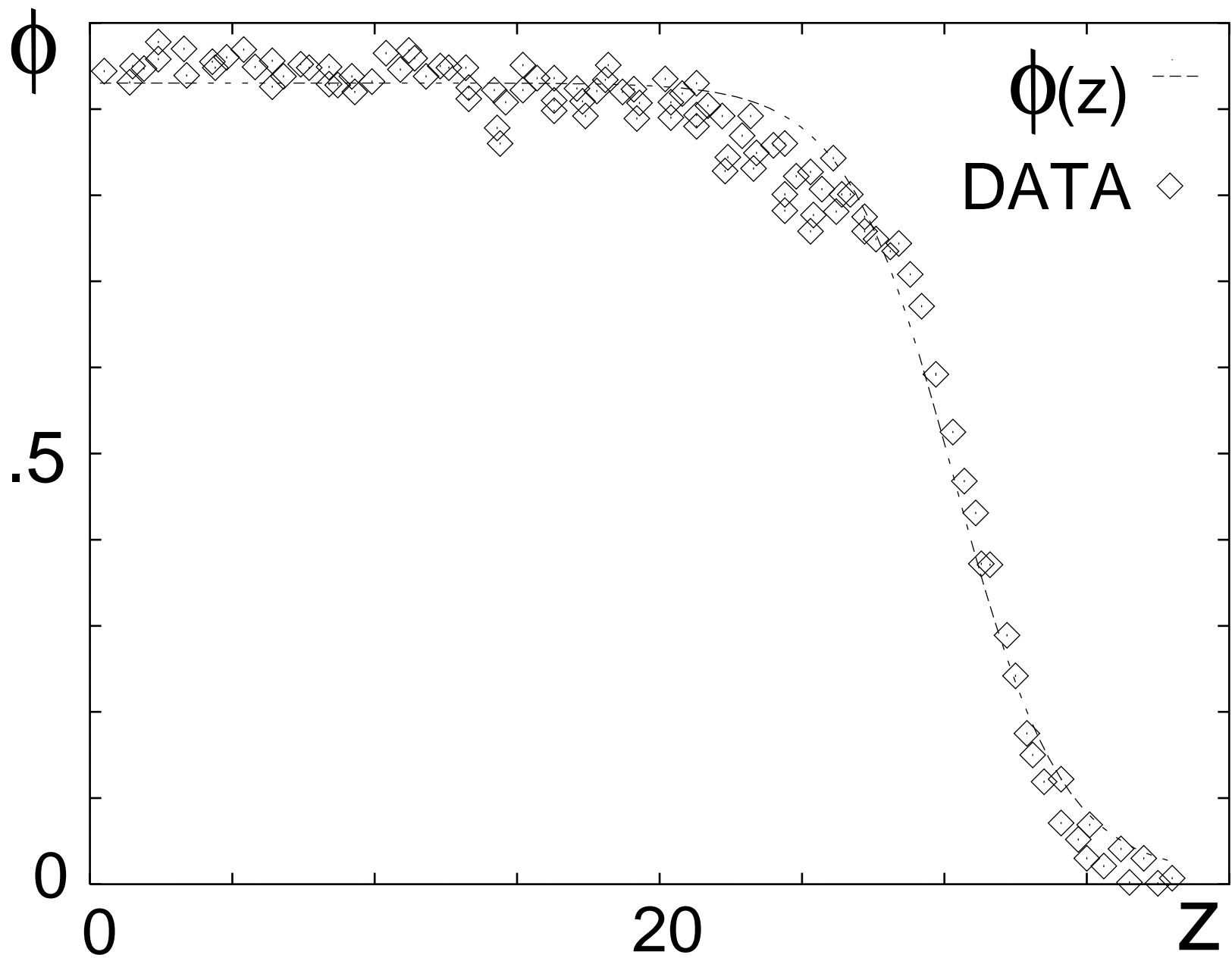
\end{document}